\documentclass{azh}
\usepackage{graphicx}

\newcommand{\km}{\,\mbox{km}\,\mbox{s}^{-1}}

\begin{document}

\title{Observations of Stellar Objects at a Shell Boundary in the
Star-Forming Complex in the Galaxy IC\,1613\thanks{
Based on observations collected with  the 6m and 1m telescopes of the Special 
Astrophysical Observatory (SAO) of the Russian Academy of Sciences (RAS), 
operated under the financial support of the Science Department of Russia 
(registration number 01-43)}
}

\author{
T.A.~Lozinskaya\inst{1}\and V.P.~Arkhipova\inst{1} \and
A.V.~Moiseev\inst{2} \and V.L.~Afanasiev\inst{2}
}

\institute{
Sternberg Astronomical Institute, Universitetski\u{\i} pr. 13,
Moscow, Russia, 119899
\and
Special Astrophysical Observatory, Nizhni\u{\i} Arkhyz, 
Karachai--Cherkessia, Russia, 369167
}

\offprints{T.A.~Lozinskaya, \email{lozinsk@sai.msu.ru}}
\date{Received May 7, 2001 }

\titlerunning{Observations of Stellar Objects in IC 1613}
\authorrunning{ Lozinskaya et al.}

\abstract{
The single region of ongoing star formation in the galaxy IC\,1613 has
been observed in order to reveal the nature of compact emission-line
objects at the edges of two shells in the complex, identified earlier in
H$\alpha$ line images. The continuum images show these compact objects
to be stars. Detailed spectroscopic observations of these stars and
the surrounding nebulae were carried out with an integral field spectrograph 
MPFS mounted on the 6m telescope of the Special Astrophysical Observatory.
The resulting stellar spectra were used to determine the spectral types
and luminosity classes of the objects. An Of star  we identified is the only 
object of this spectral type in IC\,1613. The results of optical
observations of the multi-shell complex are compared to 21cm radio
observations. The shells harboring the stars at their boundaries
constitute the most active part of the star-forming region. There is
evidence that shocks have played an important role in the formation
of the shells.
}

\maketitle

\section{INTRODUCTION}

Irregular galaxies provide unique opportunities for studies of star
formation triggered by the combined effect of stellar winds and
supernova explosions in rich stellar groupings. Due to the absence of
spiral density waves in these galaxies, the formation of giant multi-shell
complexes around stellar groupings can proceed unhindered, enabling
completion of the full ecological cycle of interaction between the stellar
and gaseous components of giant molecular clouds. For the same reason,
irregular galaxies display the longest scale lengths and time scales
on which supernovae and stellar winds play a dominant role in the
formation of new-generation stars.

IC\,1613, a faint dwarf galaxy of the Local Group, provides one of
the most striking examples of a giant multi-shell complex around a group
of young stellar associations. The northeastern sector of IC\,1613 harbors a
prominent giant complex of ionized shells and supershells surrounding
several dozen young stellar associations and star clusters. This complex is
the sole site of contemporary star formation in the galaxy (see~[1, 2]
and references therein). The complex also includes the only known
supernova remnant in the galaxy (see~[3] and references therein).
Comparison of the optical and 21cm brightness distributions showed that
the shells of HII are surrounded by extended shells of HI [4]. The shells
of ionized and neutral gas are close to each other, and partially overlap
in the plane of the sky. If the sizes of the shells along the line of
sight and in the plane of the sky are comparable, this indicates that
these shells and supershells are in physical contact with each other.
It is currently thought that collisions of massive expanding shells
with each other and/or with giant molecular clouds can serve as triggers
of gravitational instability and fragmentation in the collision region,
leading to the formation of new-generation stars.

Narrow-band H$\alpha$ images of the region considered taken in 1995 with
the 4m telescope of the Kitt Peak National Observatory (KPNO) revealed chains
of bright, compact emission-line objects located exactly at the edge
of the two shells in the complex. The same region hosts association
\#17 from the list of Hodges [5] (the eastern part of this feature was
later catalogued as association \#25 in the list [2]). However, in our
analysis in 1995, we were not able to establish a one-to-one relationship
between compact objects and stars. Moreover, the fact that the stellar
objects are located exactly along a thin shell-like structure is of
considerable interest on its own. This work was motivated primarily by
the desire to study the emission spectra and nature of the chain of
objects and their possible relationship to triggered star formation.

With this aim in view, we performed photometry and spectroscopy of this
region. Section 2 describes the instruments, observations, and data
reduction technique employed. Section 3 describes the overall structure of
the multi-shell complex derived from optical and (21cm line) radio
observations, and identifies regions that are of greatest interest for
detailed spectroscopy. Based on our photometric observations, we show
that the compact emission objects in question are stars.  Section 4
reports the results of spectroscopic observations of three selected fields
made with the integral field spectrograph MPFS mounted on the 6m telescope of 
the SAO  RAS. We
obtained the spectra of individual stars located at the shell boundaries
and estimated their spectral types and luminosity classes. We have
determined the distribution of the intensity ratios of the principal lines
in the spectra of the surrounding gaseous nebulae, and constructed the
gas radial-velocity field based on H$\beta$ and [SII] line measurements.
Section 5 discusses our main results and conclusions.

\section{OBSERVATIONS AND DATA REDUCTION}

\subsection{Photometry}

The shell complex was imaged in two filters on October 4--5, 2000
with the 1m Zeiss-1000 telescope of the Special Astrophysical
Observatory of the Russian Academy of Sciences in the process of
testing the new focal reducer SCORPIO. A description of the reducer
and transmission curves of the interference filters used can be found
at \small{\texttt{http://www.sao.ru/~moisav/scor\-pio/scorpio.html}}.
The reducer is mounted at the Cassegrain focus of the telescope
($F/13$), and the total focal ratio of the system was $F/9$.  The
spectrograph uses a TK1024 $1024\times1024$ CCD as a detector. The
system had an angular resolution of $0.52''$/pixel and a field of view
of $8.'9$.  We obtained images in two meddle-band interference
filters. The filter with a central wavelength of $\lambda_c=6620$ \AA~
and a passband halfwidth of $\Delta\lambda=190$ \AA~coadded the emission
in the H$\alpha$ and [NII] lines and with the continuum.

To obtain continuum images, we used a filter centered on
$\lambda_c=6060$~\AA~ with a passband halfwidth of $\Delta\lambda=170$~\AA.
The total exposure in each filter and the seeing during the observations
were 1800~s and $1.5''$ respectively.

\begin{table}[t!]
\caption{Log of MPFS spectroscopic observations}
\begin{tabular}{|l|c|c|c|c|}
\hline 
Field&
\multicolumn{2}{c|}{Center coordinates}& 
$T_{\exp}$, s& $z$\\
 \cline{2-3}
    & R.A. (2000.0) & DEC (2000.0)    &   &  \\
\hline
Field I&$01^{h}05^{m}6.1^s$& $+02^{\circ}09'34''$& 2700 & $42^{\circ}$ \\
Field II&$01^{h}05^{m}5.2^s$& $+02^{\circ}09'47''$& 2700 & $43^{\circ}$ \\
Field III&$01^{h}05^{m}1.8^s$& $+02^{\circ}09'35''$& 2700 & $48^{\circ}$ \\
\hline
\end{tabular}
\end{table}

After performing standard procedures for CCD frame reduction (bias
frame subtraction, flat-field correction, cosmic-hits removal), we reduced
the images to an absolute energy scale using images of the spectrophotometric
standard star BD+25$^\circ$4655 taken on the same night. The reduction
to equatorial coordinates was based on field stars, whose positions were
adopted from the digital version of the Palomar Sky Survey
\small{{\texttt{(http://stdatu.stsci.edu/dss/dss-form.html)}}.

\subsection{Integral Field Spectroscopy}

The spectroscopic observations were made on October 23--24, 2000, using
the Multipupil Field Spectrograph (MPFS) mounted at the primary
focus of the 6m telescope of SAO RAS. A
description of the spectroscope can be found at \\
\small{\texttt{http://www.sao.ru/\~\,gafan/de\-vi\-ces/mpfs/mpfs\_main.htm}}.

The new spectrograph has a larger field of view, wider spectral range, and
higher quantum efficiency than the earlier version of the MPFS [6]. The
spectrograph uses a TK1024 $1024\times1024$ CCD as a detector and enables
the spectra of 240 spatial elements (in the form of square lenses)
to be taken simultaneously, forming a $16\times15$ array on the sky.
The angular size of an image element was $1''$. A spectrum of the night-sky
background $4.'5$ from the center of the field of view is taken
simultaneously. We obtained spectra with a resolution of 8 \AA~in the
range $4350-6850$ \AA. The seeing was about $2''$.  We observed three
areas in the region of the giant shells, whose positions are shown in Fig.~2.
Table~1 gives the equatorial coordinates of the field centers, the total
exposure times T$_{exp}$, and the mean zenith angles $z$ at the time of
observation.

The spectroscopic observations were reduced using IDL-based software
developed at the SAO Laboratory of Specroscopy and Photometry of
Extragalactic Objects. The preliminary data reduction included bias frame
subtraction, flat-field correction, cosmic-hits removal, extraction of
the individual spectra from the CCD images, and wavelength calibration using
the spectrum of a He-Ne-Ar calibration lamp.

\begin{figure*}[t!]
\centering
\includegraphics[width=16 cm]{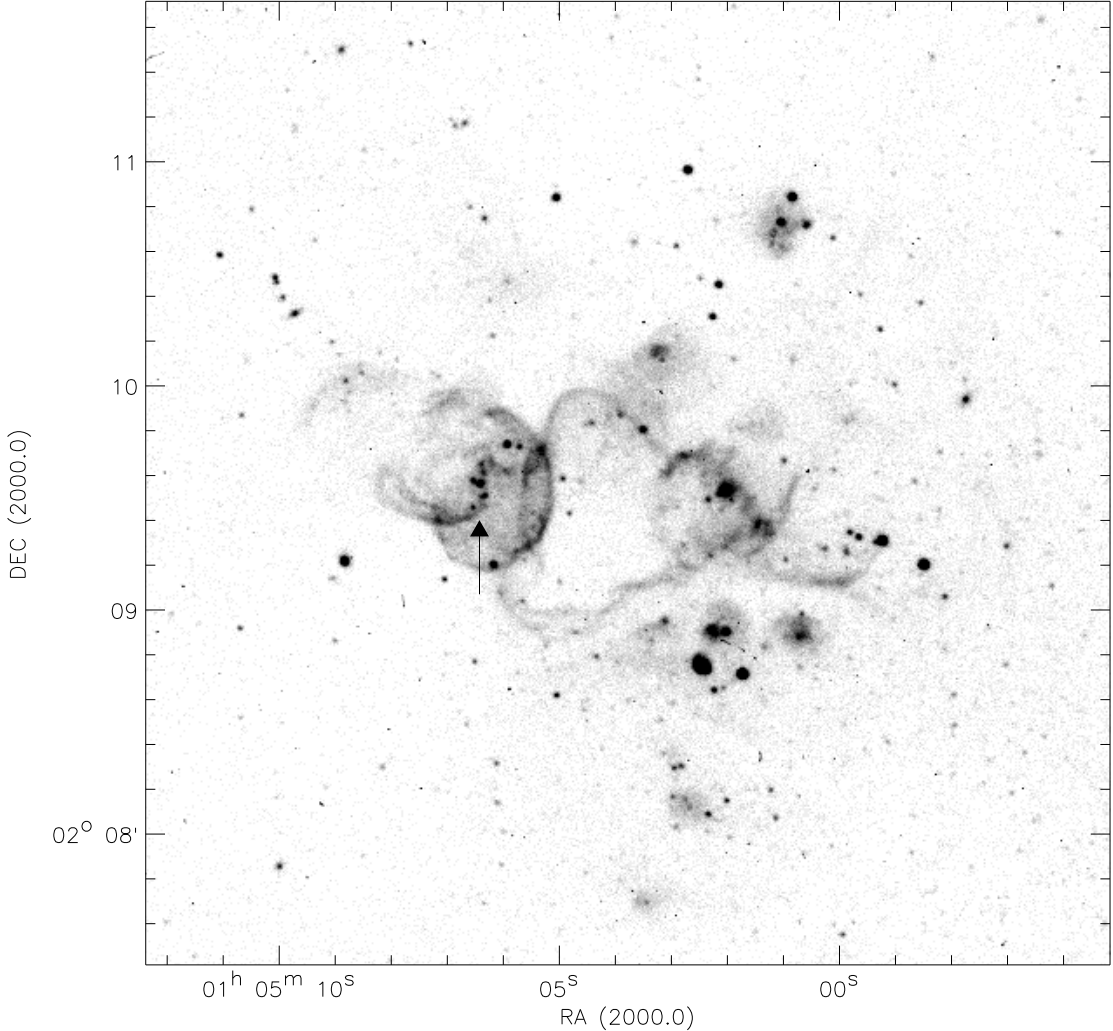}
\caption{%
(a) H$\alpha$ image of the multi-shell complex taken with the 4m
telescope of the Kitt Peak National Observatory. The arrow shows the
chain of bright, compact emission objects at the shell boundary.
(b, see color figure) HI brightness distribution (blue) superimposed on a
narrow-band H$\alpha$ image of the same area (yellow) for
the northeastern sector of the galaxy IC\,1613 (fragment of a chart
published in [4]).}%
\end{figure*}

We then subtracted the night-sky spectrum from the linearized spectra
and converted the observed fluxes to an absolute energy scale using
observations of the spectrophotometric standard star Feige 110 (taken
immediately before observing the program objects at a zenith distance of
$ z = 50^ {\circ}$). We adopted the parameters for this standard from
the public database \\
\small{\texttt{http://www.eso.org/observing/stan\-dards/spectra}}.
Our air-mass corrections were based on the mean spectral
atmospheric-extinction curve for the Special Astrophysical Observatory
given in [7].

\begin{figure*}[t!]
\centering
\includegraphics[width=15 cm]{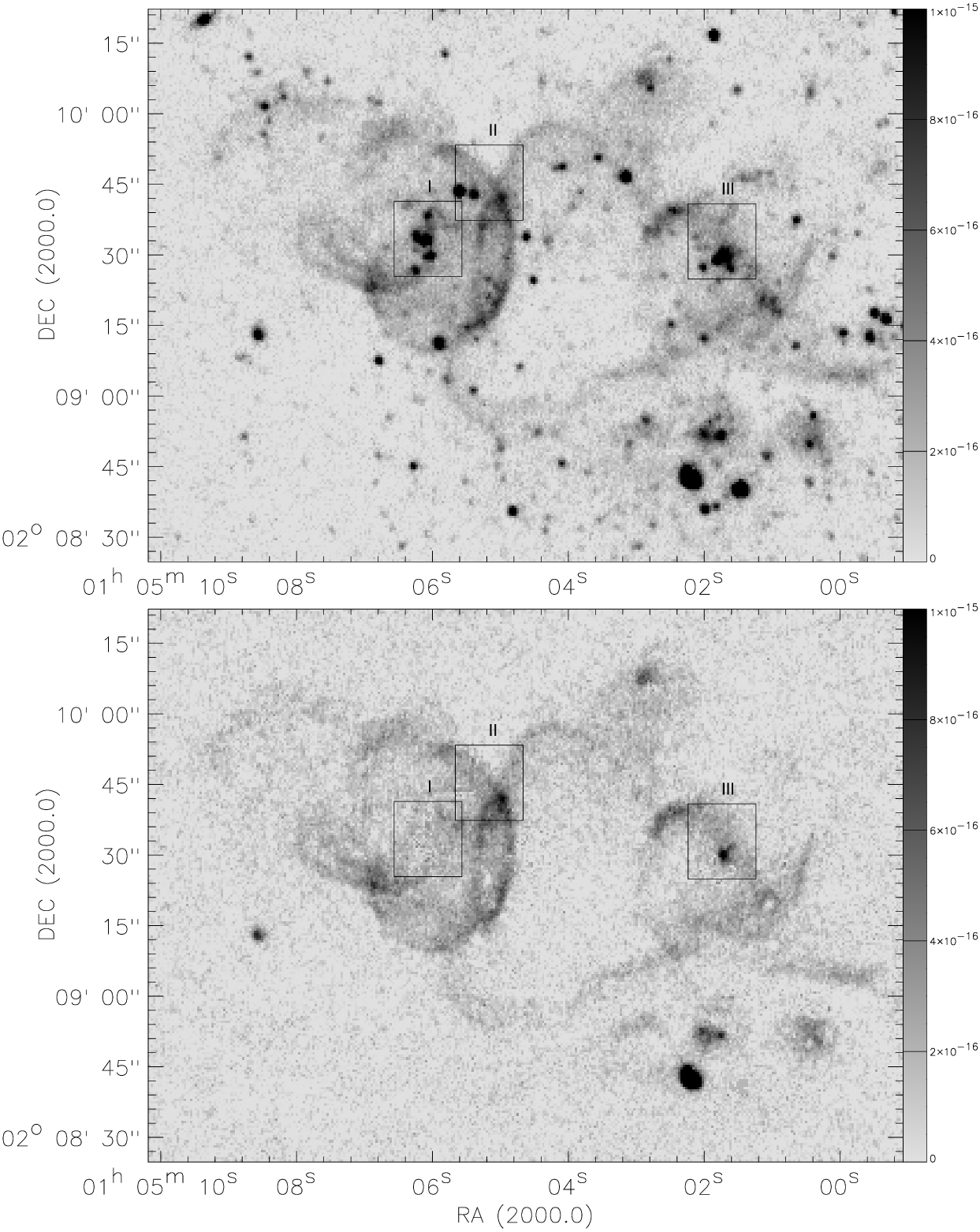}
\caption{%
Image of the complex of ionized shells taken with the Zeiss-1000
telescope equipped with the SCORPIO focal reducer. The scale is
in $erg\cdot s^{-1}\cdot cm^2/arcsec^2$. Top: image taken with a filter
centered on 6620~\AA~and a passband halfwidth of 190~\AA~coadding
emission in the H$\alpha$ and [NII] lines and the continuum. Bottom:
emission-line image (continuum subtracted). The rectangles indicate
regions observed with the SAO 6m telescope using the MPFS.}%
\end{figure*}

We compared the H$\alpha$ and [NII] line
fluxes measured using the MPFS data with those derived from images
of star-free shell regions taken with filters. The two methods
(spectrophotometric and photometric) yielded line fluxes that agreed
to within about $20\%$, providing an independent estimate of the actual
accuracy of our observations.

\section{OVERALL STRUCTURE OF THE MULTI-SHELL COMPLEX
IN THE STAR-FORMING REGION}

The multi-shell complex in the northeastern sector of IC\,1613 is the most
prominent structure seen in narrow-band H$\alpha$ images of the galaxy
[8--12]. This is the place where the overwhelming majority of the galaxy's
bright HII regions, shells, and supershells are concentrated. The stellar
component of the complex is represented by about twenty stellar associations
and star clusters [2, 5].

Figure~1a shows an H$\alpha$-line image of the multi-shell complex taken
in 1995 with the KPNO 4m telescope. The chain of bright, compact
emission-line objects located exactly at the bright rim of the shell
(indicated by an arrow) and several compact objects at the edges of
other shells are clearly visible.

The two shells with compact objects at their boundaries correspond
to objects R1 and R2 in the list of shells [12]. Both shells are located
within square N27 in Fig.~3 of [12], and include the nebulae S10 and S13
in the classification of Sandage [13].

Meaburn et al. [1] were the first to analyze the kinematics of the
complex of ionized shells. Their five spectrograms densely covered the
bright part of the complex, enabling determination of the characteristic
shell expansion velocities, which proved to be about $30\km$.
Valdez-Gutierrez et al. [12] constructed the line-of-sight velocity
field of the entire complex in the H$\alpha$ and [SII] lines and estimated
the expansion velocities of its constituent shells and supershells. HI
observations of IC\,1613 [14] showed that the complex is located in the
region of the brightest ``spot'' in the 21cm radio emission.

Lozinskaya et al. [4] were the first to make a detailed comparison
of the H$\alpha$ and 21cm radio brightness distributions of the complex.
Figure~1b shows a fragment of the map of the northeastern sector of the
galaxy published in [4], with the HI brightness distribution superimposed on
the H$\alpha$ line image shown in Fig.~1a.

An HI map with a high angular resolution of
$7.^{\prime\prime}4\times7.^{\prime\prime}0$ (corresponding to a linear
resolution of $\simeq$25~pc) was constructed using 21cm VLA observations
(part of a large project to analyze the structure and kinematics of
the neutral gas in IC\,1613 [15]).

We identified the chain of compact emission-line objects and a number of
other compact objects at the edges of the shells shown in Fig.~1a
in our H$\alpha$ line image. These compact emission-line features
could be either stars located inside compact HII regions or dense clumps
of gaseous shell material at the initial stage of gravitational
fragmentaion and/or shock-induced compression. To elucidate the nature
of these compact objects, we performed photometry of the multi-shell
complex using the Zeiss-1000 telescope equipped with the SCORPIO focal
reducer.

The upper part of Fig. 2 shows the image of the complex obtained through a
filter centered at 6620 \AA~with a halfwidth of 190 \AA. This image coadds
the emission in the H$\alpha$ and [NII] lines and the continuum.
The lower part of Fig.~2 shows the same image after subtraction of the
stellar continuum. To determine the continuum level, we used a filter with
maximum sensitivity at 6060 \AA~and a bandpass halfwidth of 170 \AA; the
shape of the filter transmission curve was rectangular rather than Gaussian.
We chose the coefficient for continuum subtraction to ensure the best
subtraction of foreground stars. Since the central wavelengths of the two
filters differ by almost 600\AA, the fact that the slope of the continuum
differs from star to star becomes important. This is why many young stars
inside the shell complex appear ``oversubtracted,'' due to the appreciable
slopes of their continuum (see Section 4). The corresponding locations are
masked in Fig.~2.

A comparison of Figs.~1 and 2 suggests that the compact objects can
be neither purely gaseous clumps nor stars located inside compact HII
regions.

The squares in Fig.~2 indicate Fields~I, II, and III for which the
spectroscopic observations reported in the next section were made.
Figures~3, 4, and 5 show enlargened images of Fields~I, II, and III,
respectively, obtained with the  SCORPIO focal reducer with a filter
centered on 6060 \AA~and a passband halfwidth of 170 \AA. These continuum
images of the three fields indicate that the compact, emission-line objects
in question are stars.

We can see 13 stars in Field~I, two of which are at the edge of the field.
The stars in Field~I belong to association \#25 in the list of [2] (the
eastern part of association \#17 from [5]). The brightest of these stars
are, indeed, located along the bright rim of the shell. We can also see
two star-like objects at the boundary of the ionized shell in Field~II. Part
of association \#17 from [5] is located in this same place. Field~III
contains stars from association \#13 from the list of Hodge (1978)
(\#18 in [2]).

Below, we report the results of spectroscopic observations of the brightest
of these stars made with a  spectrograph MPFS.

\section{RESULTS OF SPECTROSCOPIC OBSERVATIONS}

\subsection{Analysis of Integral Field Spectroscopy}

We fitted Gaussians to emission-line profiles to construct a series of
monochromatic images of the fields in the H$\alpha$ and H$\beta$
lines and in the [OIII] 5007 \AA~and [SII]$6717/6731$~\AA~forbidden lines.
All the line profiles can be adequately fitted by a single Gaussian,
without any systematic deviations. Note, however, that our spectral
resolution ($350-450\km$) substantially exceeds the expected velocities
of relative gas motions [12]. In addition, we also constructed images
of the program fields in the stellar continuum at wavelengths
$4600-4800$ \AA. Comparison of these data with the
continuum images obtained with the Zeiss-1000 telescope enabled us to coadd
the spectra from the spatial elements corresponding to individual stars.

\begin{figure*}[t!]
\centering
\includegraphics[width=15 cm]{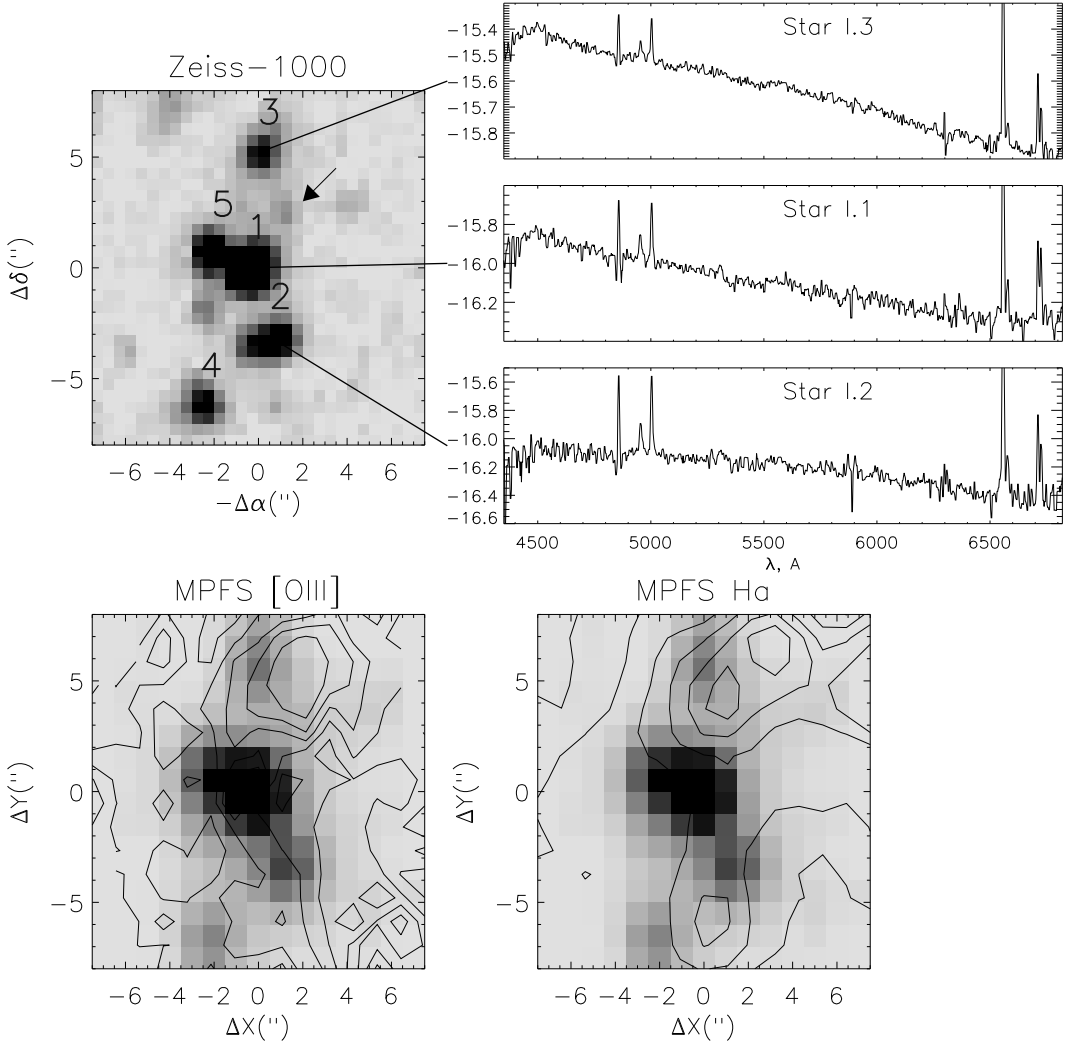}
\caption{%
Top left: Continuum image of Field 1 taken with the Zeiss-1000 telescope
through a filter centered on $\lambda_c=6060$~\AA~with a passband
halfwidth of $\Delta\lambda=170$~\AA. The star numbers are indicated;
the arrow marks the hot star that cannot be seen in the MPFS data
(see text). Top right: Spectra of the three brightest stars in the field.
Bottom: Isophotes of [OIII] (left) and $H\alpha$ (right) line images
superimposed on the $4600-4800$~\AA~continuum images (MPFS data).}%
\end{figure*}

\begin{figure*}[t!]
\centering
\includegraphics[width=15 cm]{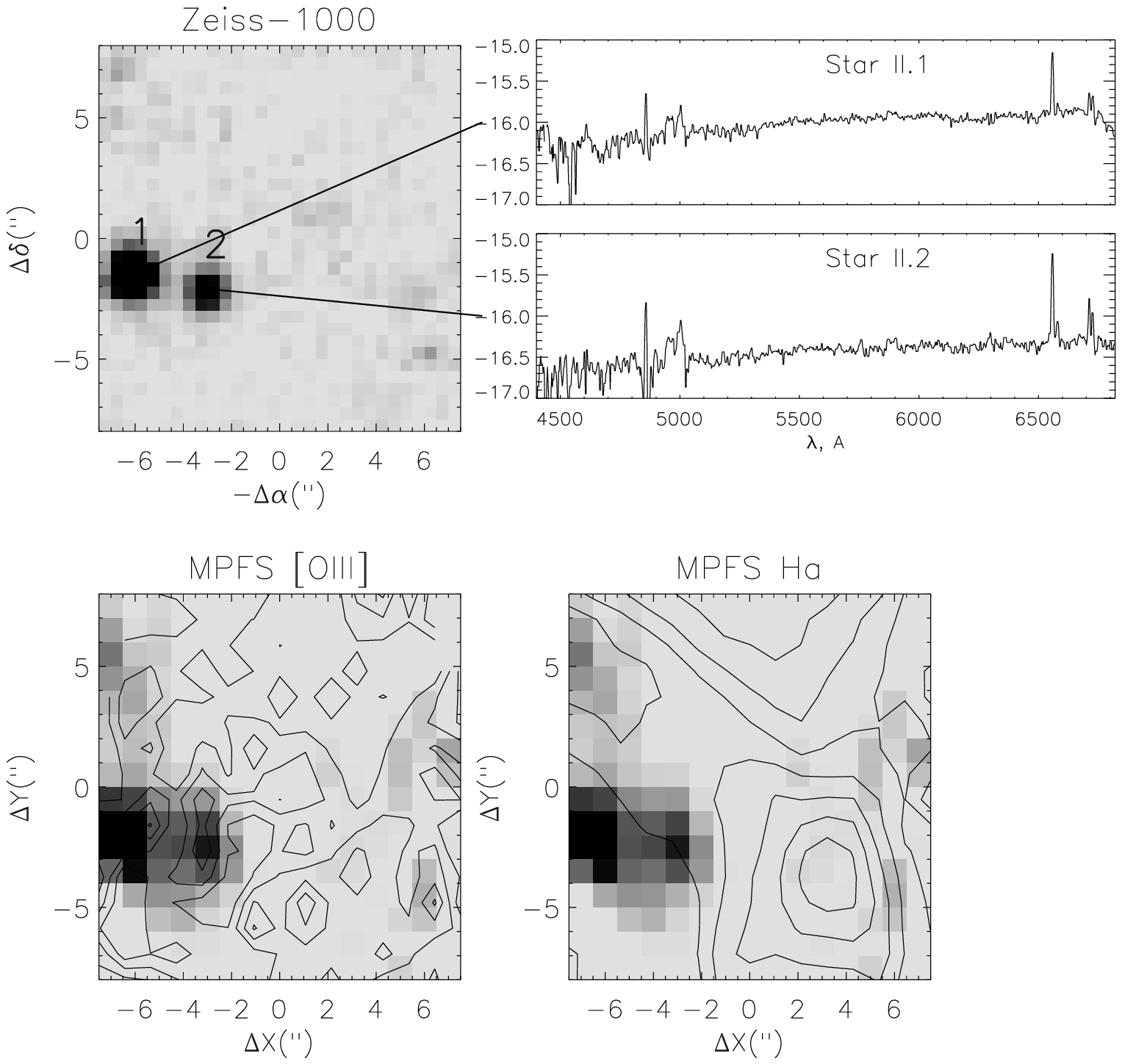}
\caption{%
Same as Fig.~3 for Field II.}%
\end{figure*}

\begin{figure*}[t!]
\centering
\includegraphics[width=15 cm]{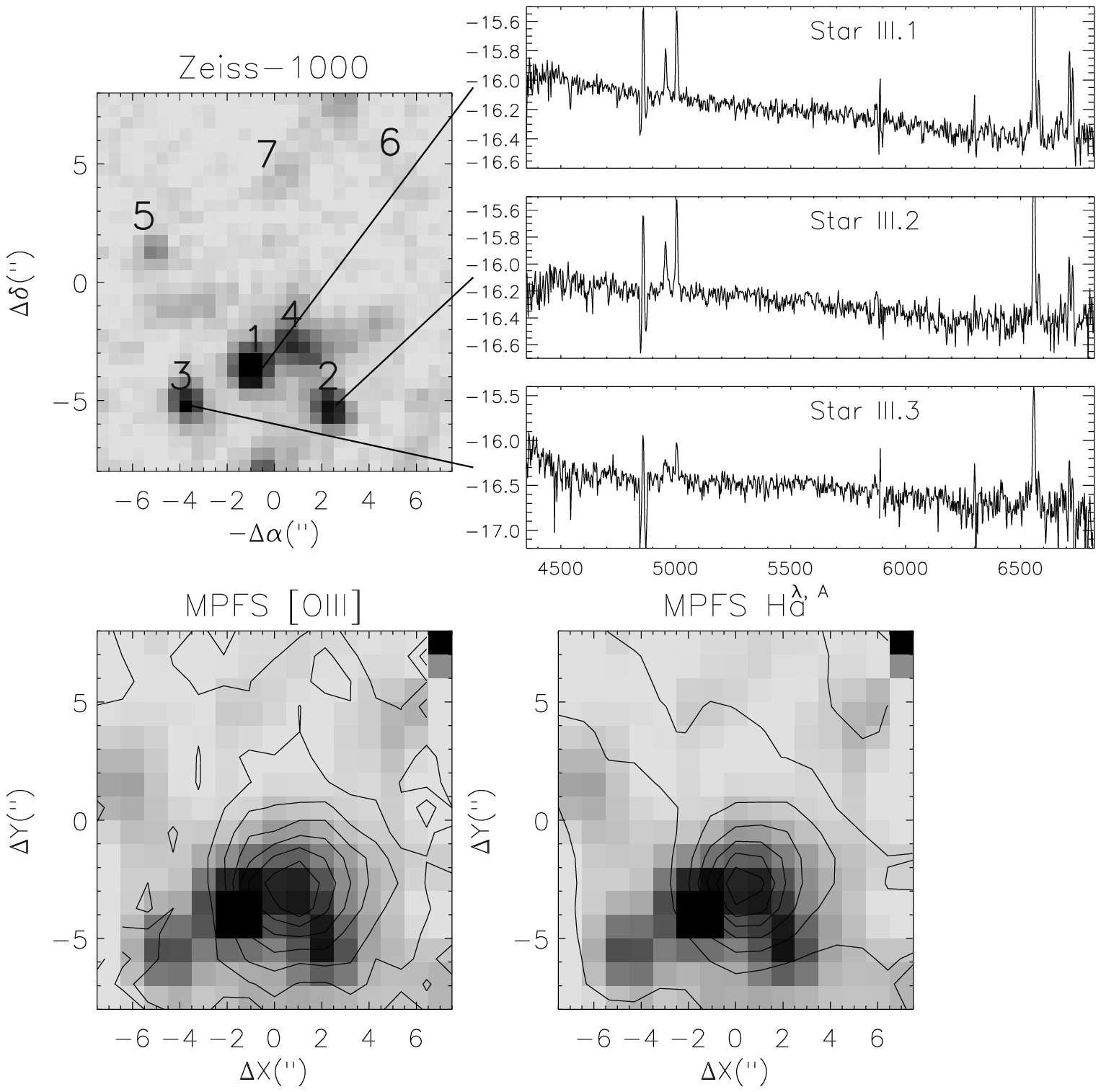}
\caption{%
Same as Fig.~3 for Field III.}%
\end{figure*}

Figures~3, 4, and 5 show the results obtained for Fields I, II, and III,
respectively (continuum and emission-line field images and the spectra
of individual stars with their surrounding nebulae).

\subsubsection{ Field I}

We analyzed the spectra of five stars in Field I to estimate their spectral
types. Unfortunately, the low resolution and fairly low signal-to-noise
ratio prevented us from reliably detecting all spectral features. In
addition, superimposed emission lines from ionized gas in the vicinity
of the stars can be seen in all the spectra.

Star I.1 can be classified as an O supergiant, judging from the HeII
4541, 4686 \AA~and SiIII 4552 \AA~lines in its spectrum.

Star I.2 is evidently much hotter than star I.1, since it exhibits
well-defined SiIV 4631, 4654 \AA~lines; the intensity of its HeII 4686 \AA
line indicates that the star is a giant rather than a supergiant.

No clearly detectable absorption lines could be found in the spectra of
stars  I.3 and I.4.

The spectrum of star I.5 exhibits a strong HeII 4686 \AA~line, indicating
the it is an O star of luminosity class III--V.

We estimated the $B$ and $V$ magnitudes of the stars in the $UBV$ system
from their absolute spectral energy distributions, based on the calibration
of [16]. The results are summarized in columns 1--3 of Table~2. Column 4
gives the $E(B-V)$ color excesses of the stars derived from the
H${\alpha}$/H${\beta}$ Balmer decrements of the nearest HII regions
located within Field I. Columns 5, 6, and 7 give the $(B-V)_{o}$ color
index, absolute magnitude $M_{v}$, and the spectral type estimated
from $(B-V)_{o}$, taking into account the luminosity classes estimated
directly from the spectra.

All estimates in this paper are based on the adopted distance modulus of
$(M-m)_{o}=24.31$, corresponding to a distance of 730 kpc, in accordance
with the new determination [17].

The hottest star in Field I appears to be star I.5.

The emission-line spectra of the gaseous environments of each star are
typical of HII regions. The emission-line ratios in the neighborhoods of
stars I.1, I.2, I.3, and I.4 are $I$([OIII]5007/H${\beta}$ = 0.8; the
relative intensities of the same lines for star I.5 yield a ratio of
1.05, confirming the higher temperature of this star.

Note that all five stars in Table~2 show appreciable interstellar extinction.

There is another hot star with coordinates $\Delta X=+2$, $\Delta Y=+3$
between stars I.1 and I.3. This star does not show up against the bright
nebular background emission in images of Field~I based on MPFS data. The
[OIII] 5007~\AA~and H${\alpha}$ brightness maps (see Fig. 3) show a
well-defined emission maximum at this location. The star can be seen on
the KPNO frame as a compact clump of H${\alpha}$ emission, and is clearly
visible on the Zeiss-1000 plate. (There are actually two stars between
stars 1 and 3, but only one is coincident with the emission peak). This
star has $V$ = 21.3$^{m}$; the extinction estimated from the Balmer decrement
is low and its temperature is high: the line intensities are
$I$([OIII]5007)/$I$(H${\alpha}$)$\simeq 0.3$ and
$I$([SII])/$I$(H${\alpha}$)$\simeq 0.2$. We therefore conclude that the
excitation of the HII region surrounding the star is radiative, and not
associated with shocks. The star is responsible for exciting both its
own surrounding gas and the gas in the vicinity of the cooler star I.3.

We used the Zeiss-1000 images to estimate the $V$ magnitudes of fainter
stars in Field I relative to star I.4 (for which Table~2 gives
$V$ = 19.8$^{m}$). The $V$ magnitudes of these stars are 20.9$^{m}$,
21.5$^{m}$, 21.6$^{m}$, 21.3$^{m}$, 21.5$^{m}$, and 21.8$^{m}$ in
order of decreasing brightness.

\subsubsection{Field II}

Our spectra of the two brightest stars in Field II show that these objects
are yellow supergiants. We used the same method as for Field I to compute
the parameters of these stars presented in Table~3, based on their
absolute spectral energy distributions.

The well-localized H${\alpha}$ and [SII] line emission maxima can
be seen to the west of star II.2, in the region with $\Delta X=+3$,
$\Delta Y=-4$. It is here that the main source of ionizing radiation
must be located. However, the emission spectrum of this region is by
no means typical of an HII region ionized by an O star: the relative
line intensities are I([OIII] 5007)/I(H${\alpha}) < 0.2$ and
I([SII])/I(H${\alpha}) \approx 0.35$. Such parameters are typical of
cooling gas behind a shock front.

The electron density of the gas component in this region is low,
as indicated by the [SII] intensity ratio of I(6717)/I(6731)$\simeq 1.7$,
corresponding to $ N_{e} < 100 cm^{-3}$.

Based on the $4600-4800$ \AA~continuum intensity in this region, we
estimate the upper limit of the $V$ magnitude of the possible ionizing
star to be $19.3^{m}$. This candidate exciting star can be seen at
$\Delta X = +6$, $\Delta Y=-4$ on the map obtained with the Zeiss-1000
telescope.

Note that the line-intensity ratio is $I$([SII])/$I$(H${\alpha}) \approx 0.35$
virtually throughout the whole of Field~II, as is typical of shock
excitation. The same field contains part of the shell R2.

\subsubsection{Field III}

We obtained the spectra of seven stars in Field~III. Table~4
summarizes the results of our photometric and spectral analysis for
these stars. The color indices were determined from the Balmer decrements
of HII regions surrounding the stars.

The spectrum of star III.1 appears to show the HeII 4541 \AA~line,
whereas the HeII 4686 \AA~line cannot be identified. We conclude that
this star is not an O supergiant, but instead a star with luminosity
class III. We cannot, however, rule out the possibility that this might
be an early O main-sequence star.

The spectrum of star III.2 exhibits a strong [OIII] 5007 \AA~nebular line:
$I$([OIII])/$I$(H${\alpha})\simeq 0.4$. This may indicate a high temperature
for the exciting star ($T\ge 3\times 10^{4}$ K). On the other hand, the
spectrum of star III.2 may contain HeI 4713 \AA~line absorption, as is
typical of early B stars. This star could well be binary.

The spectrum of star III.3 shows a well-defined emission band at
4640--4650 \AA, typical of WR and Of stars. However, the absence of other
lines characteristic of WR stars leads us to conclude that III.3 should
be classified as an Of star.

The spectrum of star III.4 shows well-defined HeII 4541 and 4686 \AA~lines,
and possibly SiIV 4630 and 4656 \AA~lines as well. We classify III.4 a an
O7--O8 star of luminosity class III.

Following are the only conclusions we can draw about the three fainter
stars in Table~4.

Star III.5 must be blue: its spectrum exhibits well-defined
traces of an He 4686~\AA~absorption line, and its estimated
$B-V$ color index is indicative of a high temperature.

Stars III.6 and III.7 must be red, judging from their spectral energy
distributions.

The main sources of gas excitation in Field III are stars III.1 and
III.4, and probably also III.2 and III.3. The H${\alpha}$ line emission
is concentrated primarily between stars III.1 and III.4, closer to
III.4. The [OIII] line emission surrounds star III.4, suggesting that
this is the hottest star in the region. According to our estimates, the
size of the [OIII] emission region is no less than $6''$, corresponding
to a linear radius of $R=10$~pc. The radius of the H${\alpha}$ emission-line
region is also about 10~pc.

The [SII] line intensity ratio in this region is
$I$(6717)/$I$(6731)$\simeq 1.5$, corresponding to $N_{e}\le$ 100 cm$^{-3}$.

We thus conclude that all four bright stars in Field III are O stars,
without a single supergiant among them. In contrast to Field I, all
stars in Field III are giants according to their luminosity classes.

\subsection{Ionized Gas Velocities}

Our spectroscopic observations in the H$\alpha$, [SII], and [OIII]
emission lines can be used to construct the line-of-sight velocity fields
of the ionized gas in Fields I, II and III. The spectral resolution of
our observations is $350-450\km$, much lower than that of
interferometric observations [12]; we therefore do not discuss the
results of our velocity measurements in detail here, pointing out only
our main conclusions. We found systematic deviations of the order of
$20-40\km$ from the mean value of $V(\mathrm{Hel})\sim -246\km$ in
each of the three fields. The velocities based on the hydrogen, sulfur,
and oxygen line measurements made in the same region vary within the
same range. The velocity variations within the regions studied do
not go beyond the interval $V(\mathrm{Hel})\simeq -200$ to $-300\km$.
(Note that, here, we refer to the velocities of the line maxima, not
those of weak features).

All the velocity variations in Fields I, II, and III indicated above
are in full agreement with the results of interferometric observations
of the complex [12].

\section{CONCLUSIONS}

We obtained new observations of the only complex with ongoing star
formation in the galaxy IC\,1613 to elucidate the nature of the chain of
compact emission-line objects we had identified earlier at the peripheries
of two shells of the complex in our analysis of deep H$\alpha$ images.
Our new continuum images obtained with the Zeiss-1000 telescope have
shown that these compact objects at the boundaries of two shells
are stars.

We performed integral field spectroscopy of selected fields in the region of
the complex using the MPFS spectrograph mounted on the 6m telescope
of the Special Astrophysical Observatory. We obtained spectra of the
stars forming the chains and estimated their spectral types and luminosity
classes. The stars in question are at different evolutionary stages.

\begin{table*}[t!]
\caption{Parameters of stars in Field I}
\begin{tabular}{|c|c|c|c|c|c|l|}
\hline
 Star &$V$&$B-V$&$E(B-V)$ &$(B-V)_{o}$ & $M_{v}$ &
 \multicolumn{1}{c}{ Sp}   \\
 \hline
 1&2&3&4&5&6&\multicolumn{1}{c}{ 7}\\
\hline
  I.1   & 17.94 & +0.02 & 0.33   &  --0.31     & --7.4 & OI      \\
  I.2   & 19.18 & --0.10 & 0.18   &  --0.28     & --5.7 & OIII    \\
  I.3   & 19.32 & +0.37 & 0.18   &  +0.19     & --5.5 & A7 III    \\
  I.4   & 19.80 & +0.10 & 0.18   &  --0.08     & --5.0 & B8 II     \\
  I.5   & 19.80 & --0.05 & 0.29   &  --0.30     & --5.4 & O8III or O4V \\
\hline
\end{tabular}
\end{table*}


\begin{table*}[t!]
\caption{Parameters of stars in Field II}
\begin{tabular}{|c|c|c|c|c|c|l|}
\hline
Star & $ V $   &  $B-V$  & $E(B-V)$ &$(B-V)_{o}$ & $M_{v}$&\multicolumn{1}{c}{ 
Sp}\\
\hline
  II.1 & 18.94 & +1.33 &  0.08  &  +1.25     & --5.6  &  G8--K0 II\\
  II.2 & 20.04 & +1.53 &  0.18  &  +1.35     & --4.8  &  K0--K2 II--I\\
\hline
\end{tabular}
\end{table*}


\begin{table*}[t!]
\caption{Parameters of stars in Field III}
\begin{tabular}{|c|c|c|c|c|c|l|}
\hline
Star &$V$ &  $B-V$  & $E(B-V)$ & $(B-V)_{o}$ &$M_{v}$&\multicolumn{1}{c}{ Sp}  
\\
\hline
  III.1 & 19.40 & +0.01 & (0.3)  & --0.29      & (--5.8)&  OIII      \\
  III.2 & 19.62 & +0.07 & 0.3    & --0.23      & --5.6  & O--B      \\
  III.3 & 20.17 & -0.06 & 0.3    &  --0.36     & --5.0  & Of ?     \\
  III.4 & 19.69 & +0.14 & (0.4)  &  --0.26     & --5.8  & OIII      \\
  III.5 & 21.2\phantom{0}  & <0?\phantom{90}  &        &       &       & 
possibly O \\
  III.6 & 21.24 &red&        &            &       &           \\
  III.7 & 21.07 &red&        &            &       &         \\
\hline
\end{tabular}
\end{table*}
Four of the stars forming the chain in Field I at the boundary of the
shell R1 are OB stars (with luminosity classes I, II, and III), while one
is a cooler object of spectral type A7 III. The position of this last star
deviates somewhat from the regular chain structure; we believe that this
A7 star is unlikely to be a member of the chain, and is simply projected
onto the region.

The two brightest stars in Field II lie at the boundary of the shell R2.
Our spectra show that these are yellow supergiants. The two stars do not
belong to associations identified in the galaxy.

The stars in Field III do not form any clear chain, but are coincident
with the brightest clump of emission at the boundary of shell R5. The
four brightest stars of the seven whose spectra we obtained are O stars.
The luminosity classes of all these stars indicates that they are giants,
with no supergiants among them, in contrast to what we found in Fields I
and II. We assigned a spectral type of Of to star III.3. This star, with
coordinates RA(2000)$=1^{\mathrm{h}}05^{\mathrm{m}}2.^{\mathrm{s}}2$,
DEC(2000)$=+02^{\circ}09'28.6''$, is the only star of this type identified
in IC\,1613 (its coordinates have been measured to within an accuracy of
$1''$).

One of the three fainter stars must also be of type O; the other two
appear to be cooler.

We have identified sources of ionization of the ambient gas in each of the
fields considered.

The shells R1 and R2, at whose boundaries the chain in Field I and two
stars of Field II are located, stand out among other objects of the
multi-shell complex in a number of ways.

(1) These two shells partially overlap in the sky, and form the
brightest emission-line region in the entire complex (see Fig.~1).
According to estimates made in [12], the H$\alpha$ luminosities of
the two shells are $\log L(\mathrm{H}\alpha) =37.51$ and $\log
L(\mathrm{H}\alpha) =37.77\,erg\cdot s^{-1}$, respectively. The same shells 
appear
to be the brightest in sulfur lines -- see Fig.~1 in [12] and Section~4
of this paper. We estimate the line intensity ratio in Field II to
be I([OIII] 5007)/I(H${\alpha}) < 0.2$, I([SII])/I(H${\alpha})
\approx 0.35$, typical of gas fluorescence behind a shock front.

(2) The expansion velocity of shell R2 ($50\km$) is the highest
in the entire complex; the expansion velocity of R1 is $30\km$ [12].
Note that Valdez-Gutierres et al.[12] measured both velocities
as the mean separation between the two maxima in the double-humped
integrated H${\alpha}$ line profiles of the corresponding shells.
In fact, images of the complex in hydrogen and sulfur lines in various
velocity intervals show both shells to display very complex kinematics.
The fact that the brightness of both objects is high even at the extremes
of the velocity interval covered -- $-272$ and $-177\km$ -- means that
the velocities of internal motions within the shells substantially
exceed the typical expansion velocities indicated above.  Two faint
symmetric features in the integrated line profile of shell R2 with
maxima at velocities of $-50$ and about $-370\km$ are also immediately
apparent (see Fig. 9 in [12]).

By analogy with the results that one of us (TAL) obtained by studying
the kinematics of supernova remnants and shells surrounding OB associations,
we suggest that these weaker high-velocity line features are closer to
the shock-front velocity than brighter low-velocity features. In this
case, the shock velocity in the shell should be at least $100\km$. It
is possible that we observe here the same ``two-component'' gas kinematics
as in the Galactic shell surrounding Cyg OB1, Cyg OB3, and a number of
other Galactic objects [18], namely the coexistence in a single shell
of bright emission at low velocities and weak emission at higher
velocities.

Such high internal velocities also reflect the substantial role played
by shocks in the formation of both shells at whose boundaries the
stellar chains are located.

(3) Comparison of the 21cm and H$\alpha$-line brightness observations
in Fig.~1b indicates that the part of the ionized shell R1 where the
stellar chain is located falls in the thinnest neutral-gas ``bridge''
connecting the two neutral shell-like structures. The characteristic
morphology of the ionized and neutral gas shells indicates that they
are in physical contact: the neutral bridge is closely pressed up against
the ionized shell R1 from outside in the region of the stellar chain,
and the two shells in this region have the same radii of curvature.
This morphology could be the result of the expanding shell R1 colliding
with the neutral bridge located between the two HI shells.

All the facts listed above suggest that the shells R1 and R2 are
located in the most dynamically active part of the star-forming
complex. This is indeed where new-generation stars are most likely
to be born. We plan further detailed high spatial- and spectral-resolution
studies of the kinematics of the neutral and ionized gas components
in this region.

\begin{acknowledgements}
This work was supported by the Russian Foundation for Basic Research
(project code 01-02-16118) and the Astronomy State Science and
Technology Program (project 1.3.1.2).  We are grateful to E. Blanton
for providing an H$\alpha$-line image of IC\,1613 taken upon our
request with the 4m telescope of the Kitt Peak National Observatory,
to V.~N.~Komarova for his assistance with the observations at the
Zeiss-1000 telescope, and to the Program Committee of the 6m telescope
for providing observing time.
\end{acknowledgements}

{\it Translated by A. Dambis}
\end{document}